\documentclass[11pt,a4paper, linktocpage,colorlinks=true, linkcolor=bluscuro,urlcolor=bluscuro,citecolor=bluscuro]{article}

\usepackage{feynmp-auto}
\usepackage{jheppub}
\usepackage{graphicx}
\usepackage{bm}
\usepackage{xcolor}
\usepackage{booktabs}
\PassOptionsToPackage{breaklinks}{hyperref}
\definecolor{bluscuro}{rgb}{0.15, 0.2, .85}
\usepackage{amsmath, amssymb}
\usepackage{slashed}
\usepackage{latexsym}
\usepackage{braket}
\usepackage[export]{adjustbox}
\usepackage{mathrsfs}
\usepackage{color}
\usepackage{multirow}
\usepackage{xspace}
\usepackage{comment}
\usepackage[caption=false]{subfig}
\usepackage{cancel}
\usepackage{url}
\usepackage{nicematrix}
\usepackage{physics}
\usepackage{pdfpages}
\usepackage{float}
\usepackage{mathtools}
\usepackage{ytableau}

\title{Gravitational multipoles from scattering amplitudes \\ in higher dimensions}

\author[a, b]{Francesco Campanella,}
\author[b]{Fabio Riccioni}

\affiliation[a]{Dipartimento di Fisica,  Universit\`a di Roma ``La Sapienza"}
\affiliation[b]{Sezione INFN Roma1, Piazzale Aldo Moro, 00184, Roma, Italy}
\emailAdd{Francesco.Campanella@uniroma1.it Fabio.Riccioni@roma1.infn.it}
\abstract{We investigate the gravitational multipole structure derived from scattering amplitudes in both four- and higher-dimensional spacetimes, with particular focus on the five-dimensional case. We develop a systematic procedure to extract multipole data from scattering amplitudes in arbitrary dimensions.
In four dimensions, only two independent multipole moments exist: mass and current moments. In this setting, we analyze the coupling of massive spin-1 and spin-3/2 fields to gravity, showing how the quadrupole and octupole structure of the Kerr solution arises from minimally coupled theories. We then extend the analysis to include non-minimal couplings, deriving the most general rotating solution with spin-induced multipoles up to octupole order.
In higher dimensions, an additional infinite family of ``stress'' multipole moments arises. Focusing on the five-dimensional case, we consider both a massive vector and a massive antisymmetric tensor coupled to gravity, and show that the resulting quadrupolar structure is qualitatively different: while the vector field produces only a mass quadrupole, the antisymmetric tensor generates only a stress quadrupole. By computing the corresponding stress-energy tensor, we demonstrate that minimally coupled theories fail to reproduce the multipolar structure of the Myers-Perry solution. This provides a direct manifestation of the breakdown of spin universality in higher dimensions.}

\begin{document}
\maketitle

\section{Introduction}

Black hole (BH) solutions are among the most fascinating and mysterious outcomes of General Relativity (GR). Their study has played a crucial role in advancing our understanding of gravitational interactions. In particular, the Kerr solution \cite{Kerr:1963ud}, which describes the metric surrounding a rotating, uncharged BH in four dimensions, is of fundamental importance for the theoretical modeling and observational interpretation of gravitational-wave signals. In this context, the multipolar structure of the Kerr geometry provides a distinctive signature, enabling precision tests of the black hole paradigm and offering a framework to constrain possible deviations arising from black hole mimickers.

While in four dimensions the Kerr metric is the unique\footnote{In this paper we do not consider solutions possessing additional charges.} rotating BH solution in GR, such uniqueness is lost in higher dimensions. Although Myers-Perry BHs \cite{Myers:1986un}, that are solutions whose horizon has spherical topology, provide the natural generalization of Kerr, in higher dimensions there exist solutions with different topologies, such as the five-dimensional black ring \cite{Emparan:2001wn} and black saturn \cite{Elvang:2007rd} solutions. Moreover, all these solutions can suffer from different classes of instabilities \cite{Emparan:2003sy,Elvang:2006dd}, signaling a complex and still only partially understood phase diagram.

An additional crucial difference between four- and higher-dimensional solutions lies in their multipolar structure. While it is well known \cite{Geroch:1970cc,Geroch:1970cd,Hansen:1974zz,Thorne:1980ru} that four-dimensional asymptotically flat vacuum solutions are characterized by two sets of multipole moments—mass multipoles and current multipoles—in higher dimensions an additional set of multipoles can arise. These have been dubbed ``stress multipoles'' in \cite{Gambino:2024uge} as they are intrinsically associated with deformations of the spatial part of the metric.

Over the past decade, scattering amplitude techniques have been extensively employed to study rotating black holes \cite{Vaidya:2014kza,Vines:2017hyw,Arkani-Hamed:2017jhn,Chung:2018kqs,Guevara:2018wpp,Guevara:2019fsj,Arkani-Hamed:2019ymq,Maybee:2019jus,Bern:2020buy,Aoude:2021oqj,Alessio:2023kgf} and the gravitational dynamics of spinning compact objects 
\cite{Kosmopoulos:2021zoq,Liu:2021zxr,Chen:2021kxt,Menezes:2022tcs,Bern:2022kto,Bautista:2023szu,Bohnenblust:2024hkw,Chen:2024mmm,Akpinar:2024meg,FebresCordero:2022jts,Jakobsen:2023ndj,Jakobsen:2022fcj,Damgaard:2022jem,Luna:2023uwd, Gonzo:2024zxo,Akpinar:2025huz,Alaverdian:2025jtw,Akpinar:2025bkt,Akpinar:2025byi,Gatica:2025uhx}. In this context, an important development has been the result of \cite{Chung:2018kqs}, based on the massive spinor-helicity formalism introduced in \cite{Arkani-Hamed:2017jhn}, showing that the 3-point amplitude describing the emission of a graviton from a minimally coupled massive particle reproduces, in the infinite-spin limit, the stress-energy tensor of a Kerr BH. A crucial ingredient in establishing this correspondence is the expression of the stress-energy tensor for Kerr in momentum space at linear order in the gravitational coupling as a function of the angular momentum \cite{Vines:2017hyw}. The minimally coupled amplitude for a particle of spin 
$s$ reproduces, in the classical limit, this stress-energy tensor expanded up to order $2s$ in the angular momentum, as first shown up to massive spin 2 in \cite{Vaidya:2014kza}.

In order to generate the multipole moments of more general rotating solutions from scattering amplitudes, it is necessary to include higher-derivative interactions \cite{Vaidya:2014kza}. Indeed, as shown in \cite{Gambino:2024uge}, introducing a higher-derivative coupling for a massive spin-1 field yields the vacuum Hartle–Thorne solution \cite{Hartle:1967he,Hartle:1968si}, which describes the spacetime of a generic spinning object up to quadratic order in the angular momentum. Furthermore, in \cite{Bianchi:2024shc} it was shown that the most general spin-induced stress-energy tensor in momentum space can be expressed in terms of two functions of the angular momentum, which directly encode the mass and current multipolar structure of a given solution through their series expansions. In the case of Kerr, these functions precisely reproduce the stress-energy tensor derived in \cite{Vines:2017hyw}.

Extending this analysis to higher dimensions, one finds that an additional function of the angular momentum appears in the stress-energy tensor, whose series expansion encodes the coefficients of the stress multipoles at all orders \cite{Bianchi:2024shc}. This result can be applied to Myers–Perry BHs \cite{Myers:1986un}, for which the corresponding source in momentum space admits a remarkably elegant expression in terms of Bessel functions, naturally generalizing the four-dimensional case. A notable consequence is that the stress-energy tensor can be Fourier-transformed to coordinate space, yielding a singular matter configuration that takes the form of a lower-dimensional distribution and reproduces the singularity structure of the fully non-linear black hole solution \cite{Bianchi:2024shc}. This provides the natural higher-dimensional generalization of the equatorial, pressureless thin disk rotating at superluminal speed that sources the Kerr solution \cite{Israel:1970kp}.

From a scattering-amplitude perspective, a natural question is what form the 3-point amplitude for graviton emission from a massive field takes in higher dimensions in order to reproduce the stress-energy tensor of Myers–Perry BHs. In \cite{Gambino:2024uge}, it was shown that, already in five dimensions, a massive vector requires the inclusion of non-minimal couplings in order to reproduce the quadrupolar structure of the Myers–Perry solution. The aim of this paper is to clarify some aspects that remained unresolved in that analysis.
In particular, it is well-known that in higher dimensions the very notion of spin must be reconsidered, as massive tensor fields in mixed-symmetry representations of the Lorentz group give rise to degrees of freedom that differ from those of totally symmetric representations. As we will show, this implies that the multipolar structure of the stress-energy tensor depends sensitively on the specific representation under consideration. In this sense, the notion of spin universality—namely, the fact that a minimally coupled theory for arbitrary spin $s$ reproduces the same multipoles up to order $2s$—is not even well defined in higher dimensions.

In order to set the stage for the higher-dimensional case, we first identify the appropriate procedure to extract spin dependence from scattering amplitudes in four dimensions. We then apply this framework to vector and spin-3/2 fields, explicitly showing how minimally coupled theories reproduce the quadrupole and octupole moments of Kerr. We also demonstrate how, in both cases, the inclusion of non-minimal couplings allows one to obtain the most general stress-energy tensor up to cubic order in the angular momentum.

We subsequently extend the analysis to five-dimensional spacetime, considering both a vector field and an antisymmetric tensor coupled to gravity. In both cases, the corresponding 3-point amplitudes yield classical contributions up to quadratic order in the angular momentum. However, while the vector case only generates a mass quadrupole, the antisymmetric tensor produces exclusively a stress quadrupole. Moreover, in the minimally coupled case, neither of these coefficients matches those of the Myers–Perry solution.

The paper is organized as follows.
In Section \ref{sez 2} we review the multipolar structure in General Relativity and the procedure to compute it from scattering amplitudes in arbitrary dimensions.
In Section \ref{sez 3} we analyze the multipolar structure in four-dimensional spacetime, considering the coupling of gravity to massive spin-1/2, spin-1 and spin-3/2 fields. 
In Section \ref{sez 4} we extend the analysis to five-dimensional spacetime, focusing on the minimal coupling of gravity to a massive vector field and a massive antisymmetric tensor.
Finally, Section \ref{sez 5} contains our conclusions.

\textbf{Conventions.} We work in mostly negative signature with $\eta_{00}=+1$ and in natural units, $\hbar=c=1$, whereas we keep the gravitational coupling constant $G$ explicit. Greek indices $\mu,\nu=0,1,...,d$   are for spacetime components and the Latin indices are for space components only $i,j=1,...,d$, where $D=d+1$ is the number of spacetime dimensions. Moreover Greek indices $\alpha,\beta,..$ are spinor indices. Repeated Latin indices are always meant to be contracted by $\delta^{ij}$, and there is no sign difference between upstairs and downstairs indices.

\section{Gravitational multipoles from scattering amplitudes}
\label{sez 2}

The multipolar expansion derived from 3-point scattering amplitudes in the spinor-helicity formalism arises from an appropriate identification of the spin tensor in terms of spinor-helicity variables. In the case of minimally coupled fields, this leads to the well-known exponentiation structure, which reproduces the multipolar expansion of the Kerr solution \cite{Chung:2018kqs}. The aim of this section is to show how the same spin tensor expansion arises from a field-theoretic perspective, through the correct identification of spin in terms of polarization tensors \cite{Vaidya:2014kza}.

\begin{figure}[ht]
    \centering
    \includegraphics[width=0.4\linewidth]{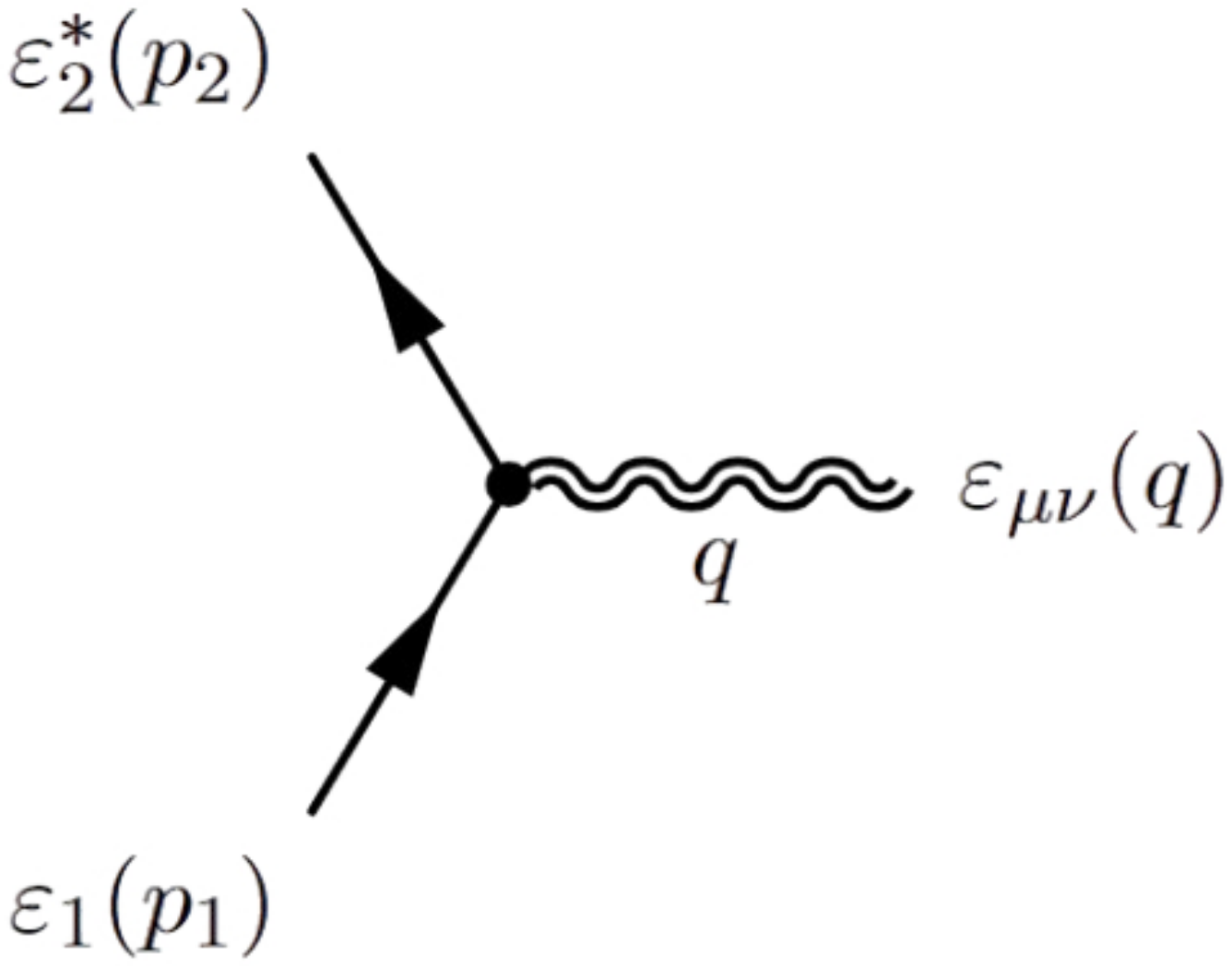}
    \caption{3-point amplitude involving spin massive fields and a graviton.}
    \label{tree}
\end{figure}

From a  field-theory perspective, one reads the stress-energy tensor from the amplitude in Figure \ref{tree}, where the
two massive particles have equal mass and spin and $q= p_1- p_2 $ is the transferred momentum. We can write the massive-particle momenta in terms of the center-of-mass momentum and the transferred momentum as
\begin{equation}
    p_1^\mu=P^\mu+\frac{q^\mu}{2}, \quad\quad p_2^\mu=P^\mu-\frac{q^\mu}{2} \ .
\label{momenti 2}
\end{equation}
The angular momentum is computed directly from the stress-energy tensor \cite{Bjerrum-Bohr:2002fji},  resulting in the identification
\begin{equation} \label{eq: SO(d)tensorfullycov}
    J^{\mu\nu}=\frac{\varepsilon^{B}_2(M^{\mu\nu})_B{}^{A}\varepsilon_1{}_{A}}{\varepsilon_2\cdot\varepsilon_1} \ ,
\end{equation}
where the indices $A$ and $B$ denote a generic, arbitrary spin, representation for the polarization tensors, and $(M^{\mu\nu})_A{}^{ B}$ are the Lorentz generators in that representation, satisfying the algebra
\begin{equation}
    [M^{\mu \nu}, M^{\rho \sigma}]=-i\Big(\eta^{\mu\rho}M^{\nu\sigma}-\eta^{\nu \rho}M^{\mu \sigma}+\eta^{\nu \sigma}M^{\mu \rho}-\eta^{\mu\sigma}M^{\nu\rho}\Big)\ .
    \label{Lorentz algebra}
\end{equation} 
In the expression above, the polarizations are the ones of the center-of-mass momentum, implying that in order to read the expansion in the spin tensor inside the amplitude, one has to boost them back by $\pm\frac{q}{2}$.
This is precisely what the Spin Supplementary Condition (SSC for short) takes care of  (see {\it e.g.} \cite{Guevara:2018wpp} and references therein). 

In the context of the analysis performed in this paper, we can choose the frame to be the center-of-mass frame, so that the tensor in eq. \eqref{eq: SO(d)tensorfullycov} becomes the $SO(d)$ tensor with only spatial indices
\begin{equation} 
    J^{ij}=\frac{\varepsilon^{B}_2(M^{ij})_B{}^{A}\varepsilon_1{}_{A}}{\varepsilon_2\cdot\varepsilon_1} \ ,
    \label{SO(d) little group}
\end{equation}
where $(M^{ij})_A{}^{B}$ are the little group generators. This object automatically satisfies SSC.
The strategy to compute the angular-momentum dependence within a 3-point scattering amplitude is therefore to expand the polarizations around the static frame in powers of the transferred momentum \cite{Vaidya:2014kza}. In the following we are going to describe the procedure in detail.   

A generic theory describing a massive field coupled to gravity yields a 3-point amplitude which is a function of the massive momenta $p_1$ and $p_2$ and the transferred momentum. We first rewrite the amplitude in terms of $P$ and $q$, and we observe that in the center-of-mass frame the first in purely time-like while the second is purely space-like.
On-shellness of the graviton polarization $\epsilon_{\mu\nu}$ implies that terms proportional to $q^\mu$  and $q^\nu$, as well as terms proportional to $\eta^{\mu\nu}$, vanish. From the point of view of the classical theory, this corresponds to neglecting terms of the metric that can be gauged away within the gauge chosen \cite{Gambino:2024uge}, and therefore do not contribute to the multipolar expansion. We further neglect terms proportional to $q^2$, that would lead to local contributions to the metric. This implies that we can substitute $P^\mu$ with $m u^\mu$, where $u^\mu=\delta^\mu_0$ is the velocity of the source.

In the amplitude, the polarization tensors $\varepsilon_{1,2}(p_{1,2})$  are also expanded around the rest frame polarizations $\varepsilon_{1,2}$,\footnote{Throughout the paper, we always specify the momentum when we refer to the polarization of the particle, while we use the same symbol without explicit dependence on the momentum when we refer to the polarization boosted back to the rest frame of the particle.} and this gives in general terms of the form $\varepsilon_{1,A} \varepsilon_2^B$ in different irreducible representations of the little group $SO(d)$. Using the specific form of the generators $(M^{ij})_B{}^A$,  each irreducible representation $\mathcal{R}_{\mathcal{I}}$  can be uniquely written in terms of specific products $\underbrace{M \cdot M \cdot \ldots \cdot M}_{\mathcal{R}_{\mathcal{I}}}$ of the generators $M$'s  projected on the same representation, as
\begin{equation} 
   \underbrace{ \varepsilon_{2}^B \varepsilon_{1,A} }_{\mathcal{R}_{\mathcal{I}}}  = C_{\mathcal{R}_{\mathcal{I}}}(\underbrace{ M \cdot M \cdot \ldots  \cdot M }_{\mathcal{R}_{\mathcal{I}}})_A{}^B \bigg[\varepsilon_2^D(\underbrace{M \cdot M \cdot \ldots \cdot M}_{\mathcal{R}_{\mathcal{I}}})_D{}^C\varepsilon_{1,C}\bigg] \ .
    \label{Jirrep}
\end{equation}
Generalizing eq. \eqref{SO(d) little group}, one then  determines the various powers of the angular momentum in the amplitude through the identification
\begin{equation}
    \underbrace{J\cdot J \cdot \ldots\cdot  J}_{\mathcal{R}_{\mathcal{I}}}=\frac{1}{\varepsilon_2\cdot\varepsilon_1}\varepsilon_2^D(\underbrace{M\cdot M \cdot \ldots \cdot  M}_{\mathcal{R}_{\mathcal{I}}})_D{}^C\varepsilon_{1,C} \ ,
\end{equation}
where the number of generators involved in the representation $\mathcal{R}_{\mathcal{I}}$ gives the order of the multipole. Note that in general not all the representations that appear in the decomposition \eqref{Jirrep} correspond to actual multipoles. By consistency, these representations must disappear in the amplitude in the classical limit by symmetrisation when contracted with the momenta up to local or gauge terms.  

Once this analysis is carried out, so that the amplitude is written in terms of of the spin tensor $J_{ij}$, one follows~\cite{Kosower:2018adc} to schematically obtain the classical limit out of scattering amplitude calculations. This is obtained making the replacement
\begin{equation}
    q\rightarrow \hbar\,  q \qquad \text{and} \qquad { J}\rightarrow\frac{1}{\hbar}{ J} \ , \label{eq:KMOC}
\end{equation}
keeping only the terms $O(\hbar^0)$, \textit{i.e.} those that survive in the limit $\hbar\rightarrow 0$. Therefore, terms $O(q^n)$ with $n>0$ are quantum, while terms $O(({ J} q)^n$) are classical. Crucially, the presence of the spin compensates the transferred momentum and gives rise to classical terms with higher powers in $q$. 

The outcome of this analysis is that if one considers a theory in which generic higher-derivative couplings are turned on, one obtains at linear order in the gravitational coupling the most general stress-energy tensor as a function of the spin and transferred momentum subject to the constrains outlined above. In \cite{Bianchi:2024shc} it was shown that such tensor, in arbitrary dimension, has the following expression:
\begin{equation}
\begin{split}
    T^{\mu\nu} (q)& =u^\mu u^\nu m\sum_{n=0}^{+\infty}F_{2n,1}\zeta^{2n} +m\sum_{n=0}^{+\infty}F_{2n+2,2}(S\cdot q)^\mu(S\cdot q)^\nu\zeta^{2n}\\&-\frac{i}{2}mq_\rho(u^\mu S^{\nu\rho}+u^\nu S^{\mu\rho})\sum_{n=0}^{+\infty}F_{2n+1,3}\zeta^{2n}
    \ ,
\end{split}
\label{EMT}
\end{equation}
where
\begin{equation}
   S = J/m
\end{equation}
is the spin per unit mass, and we have defined \begin{equation}
    \zeta\equiv\sqrt{q\cdot S\cdot S\cdot q} = \sqrt{q^\mu S_{\mu}{}^{\nu}S_{\nu}{}^{\sigma}q_{\sigma}} \ .
    \end{equation}
    The $F$'s are constant coefficients dubbed `form factors', and the universality of the mass and dipole terms implies that $F_{0,1} = F_{1,3}=1$.\footnote{The definition of $\zeta^2$ in terms of $ S $ and $q$ has opposite sign compared to \cite{Bianchi:2024shc} due to the opposite choice of the metric signature.}

In order to relate the form factors to the multipolar expansion, it is convenient to follow the strategy originally introduced by Thorne, which consists in defining the so-called Asymptotically Cartesian Mass Centered~(ACMC) coordinates, a particular reference frame in which it is possible to unambiguously extract the multipole moments directly from the asymptotic behavior of the metric~\cite{Thorne:1980ru}. Albeit not gauge invariant, Thorne's formalism is more intuitive than the gauge invariant one pioneered by  Geroch and Hansen~\cite{Geroch:1970cc, Geroch:1970cd, Hansen:1974zz}, and it can be shown that the two definitions of multipole moments coincide~\cite{Gursel:1983nkl} (see Refs.~\cite{Cardoso:2016ryw,Mayerson:2022ekj} for a review). More recently, the extension of this analysis to higher dimensions has been discussed in the Geroch-Hansen framework in~\cite{Tanabe:2010ax} and using the ACMC coordinates \`a la Thorne in~\cite{Heynen:2023sin,Gambino:2024uge}. 

We consider the multipole expansion of the linearized metric in ACMC coordinates in arbitrary dimension~\cite{Gambino:2024uge} using the following normalization
\begin{equation}\label{eq:MultipoleExpandedMetric}
\begin{aligned}
    g_{00}&=-1+4\frac{d-2}{d-1}\sum_{\ell=0}^{+\infty}\frac{G m \rho(r)}{r^\ell}\mathcal{M}^{(\ell)}_{A_\ell}N_{A_\ell}+\cdots\ ,\\
    g_{0i}&=2(d-2)\sum_{\ell=1}^{+\infty}\frac{Gm\rho(r)}{r^\ell}\mathcal{J}^{(\ell)}_{i, A_\ell}N_{A_\ell}+\cdots\ ,\\ 
    g_{ij}&=\delta_{ij}+4\frac{d-2}{d-1}\sum_{\ell=2}^{+\infty}\frac{G m \rho(r)}{r^\ell}\tilde{\mathcal{G}}^{(\ell)}_{ij,A_\ell}N_{A_\ell}+\cdots\ ,
\end{aligned}
\end{equation}
where $A_\ell=a_1\cdots a_\ell$,  $N_{A_\ell}=\frac{x_{a_1}\cdots \,x_{a_\ell}}{r^\ell}$,  $\mathcal{M}^{(\ell)}_{A_\ell}$ and $\mathcal{J}^{(\ell)}_{i, A_\ell}$ are the mass and current multipole moments, respectively, and $\tilde{\mathcal{G}}^{(\ell)}_{ij,A_\ell}$ are related to the stress multipoles through the relation 
\begin{equation}
\mathcal{G}^{(\ell)}_{ij,A_\ell}=\tilde{\mathcal{G}}^{(\ell)}_{ij,A_\ell}+\frac{1}{2}\delta_{ij}\Big(\mathcal{M}^{(\ell)}_{A_\ell}-\tilde{\mathcal{G}}^{(\ell)}_{kk,A_\ell}\Big)\ ,
\end{equation}
with $\mathcal{G}^{(\ell)}_{ij,A_\ell}$ being the stress multipoles defined in~\cite{Gambino:2024uge} (see also~\cite{Amalberti:2023ohj}). 

By inserting the stress-energy tensor in eq. \eqref{EMT} as a source for the metric, one is thus able to relate the form factors to the multipolar expansion. This results in the expressions 
\begin{equation}
\begin{split}
    &\mathcal{M}_{A_{2\ell}}^{(2\ell)}=\frac{(d+4\ell-4)!!}{(d-2)!!}(-1)^\ell \bigg(F_{2\ell,2}+(d-2)F_{2\ell,1}\bigg)(S\cdot S)_{A_{2\ell}}\bigg|_{\text{STF}} \ , \\&
    \mathcal{J}_{i,A_{2\ell+1}}^{(2\ell+1)}=\frac{(d+4\ell-2)!!}{(d-2)!!}(-1)^\ell F_{2\ell+1,3}S_{ia_1}(S\cdot S)_{A_{2l}}\bigg|_{\text{ASTF}} \ , \\&
    \mathcal{G}_{ij,A_{2\ell}}^{(2\ell)}=(d-1)\frac{(d+4\ell-4)!!}{(d-2)!!}(-1)^\ell F_{2\ell,2}S_{ia_1}S_{ja_2}(S\cdot S)_{A_{2\ell-2}}\bigg|_{\text{RSTF}} \ ,    
\end{split}
\label{multipoli}
\end{equation}
for the mass multipoles, the current multipoles and the stress multipoles respectively \cite{Bianchi:2024shc}. In these expressions we are using the short-hand notation
\begin{equation}
    (S\cdot S)^{A_{2n}}=(S\cdot S)^{a_1a_2}\cdots(S\cdot S)^{a_{n-1}a_n} \ ,
\end{equation}
and we denote  with STF, ASTF and RSTF the irreducible representations of $SO(d)$ identified by the Young Tableax
\begin{equation}
\label{STFASTFRSTF}
\begin{split}
    \text{STF}&:\ \overbrace{\begin{ytableau}
        {} & {} & {}
    \end{ytableau} \dots \begin{ytableau}
        {} & {}
    \end{ytableau}}^{2\ell} \\
    \text{ASTF}&:\ \overbrace{\begin{ytableau}
        {} & {} & {}\\
        {}
    \end{ytableau}\dots\begin{ytableau}
        {} & {} & {}
    \end{ytableau}}^{2\ell+1} \\
    \text{RSTF}&:\ \overbrace{\begin{ytableau}
        {} & {} & {}\\
        {} & {}
    \end{ytableau}\dots\begin{ytableau}
        {} & {}
    \end{ytableau}}^{2\ell} \ .
\end{split}
\end{equation}
One can show that in four dimensions, {\it i.e.} for $d=3$, the stress multipoles identically vanish, and therefore in this case a  rotating solution is characterized by the mass and current multipoles alone. Besides, using the epsilon symbol one can show that the current multipoles reduce to STF representations with $2\ell +1 $ indices. 

Our aim is to compare the multipolar expansion that one obtains at a given order from scattering amplitudes to multipoles of the 
Myers-Perry solution, describing a rotating BH in arbitrary dimension. Such multipoles where determined in \cite{Bianchi:2024shc} and they are identified by the form factors
\begin{equation}\label{eq:FFMParbitraryD}
\begin{aligned}
    F_{2\ell, 2}&=\frac{1}{2}\Omega(d) \frac{(-1)^\ell}{(\ell-1)!\ \Gamma\left(\ell+1+\frac{d-2}{2}\right)}\left(\frac{d-1}{4}\right)^{2\ell-1+\frac{d-2}{2}}\ , \\
    F_{2\ell+1, 3}&=\Omega(d) \frac{(-1)^\ell}{\ell!\ \Gamma\left(\ell+1+\frac{d-2}{2}\right)}\left(\frac{d-1}{4}\right)^{2\ell+\frac{d-2}{2}}\ ,\\
    F_{2\ell, 1}&=F_{2\ell, 2}+F_{2\ell+1, 3}\ ,
\end{aligned}
\end{equation}
where 
\begin{equation}
    \Omega(d)=\frac{\Gamma(d/2)}{2^{2-d}(d-1)^{\frac{d-2}{2}}} \ .
\end{equation}
Remarkably, by substituting the form factors in eq. \eqref{EMT}, the series expansions can be nicely resummed in terms of Bessel functions, giving the stress-energy tensor of the Myers-Perry solution
\begin{equation}
\begin{aligned}
    T^{\mu\nu}(q)\big|_{\text {MP}} & =m \Omega(d) \zeta^{-\frac{d-2}{2}}\bigg[ u^\mu u^\nu \bigg(J_{\frac{d-2}{2}}\bigg(\frac{d-1}{2}\zeta\bigg)-\frac{\zeta}{2} J_{\frac{d}{2}}\bigg(\frac{d-1}{2} \zeta\bigg)\bigg)\\&
    -\frac{1}{2\zeta}(S\cdot q)^\mu(S\cdot q)^\nu J_{\frac{d}{2}}\bigg(\frac{d-1}{2}\zeta\bigg)\\&
    -\frac{i}{2}\bigg(u^\mu(S\cdot q)^\nu+u^\nu(S\cdot q)^\mu\bigg)J_{\frac{d-2}{2}}\bigg(\frac{d-1}{2}\zeta\bigg)\bigg] 
\end{aligned}
\label{Myers-Perry}
\end{equation}
in any number $d$ of spatial dimensions. It is amusing to observe how this equation resembles the form of the metric in momentum space 
that can be exactly computed in the Kerr-Schild gauge~\cite{Bianchi:2025xol}.

A specific analysis is needed in four dimensions. Indeed, the absence of stress multipoles has a natural counterpart in momentum space, namely the fact that the terms $F_{2n,2}$ in eq. \eqref{EMT} can all be reabsorbed in $F_{2n,1}$ up to local terms or terms that can be gauged away. This can be easily seen in the center-of-mass frame writing $S^{ij} = \epsilon^{ijk} s_k$ in order to obtain
\begin{equation}
    (S\cdot q)^i (S\cdot q )^j =\delta^{ij}(\vec{s}^{\,2} \vec{q}^{\,2} -(\vec{s}\cdot \vec{q}\, )^2)+(s^i q^j+s^j q^i )\vec{s}\cdot \vec{q} - s^i s^j \vec{q}^{\,2}  - q^i q^j \vec{s}^{\,2} \ ,
    \label{Duality S tensor}
\end{equation}
and noticing that only the second term on the right-hand side should be considered. In covariant notation this can be written as
\begin{equation}
     (\eta_{\mu\nu} - u^\mu u^\nu )\vec{s}^{\,2} (\vec{q}^{\,2}- \vec{q}_\perp^{\, 2} ) \ ,
\end{equation}
where with $q_\perp$ we denote part of the transferred momentum which is orthogonal to the spin vector. Among the terms above, 
the $\eta^{\mu\nu}$ and $\vec{q}^{\,2}$ terms are pure gauge, so that one is left with $u^\mu u^\nu \vec{s}^{\,2} \vec{q}_\perp^{\, 2}$, which is equal to $u^\mu u^\nu q\cdot S \cdot S \cdot q$ as expected. As a consequence, 
the most general stress-energy tensor for a vacuum solution with a given spin-induced multipolar structure  can be written as
\begin{equation}
    T^{\mu\nu} (q)\big|_{d=3} =u^\mu u^\nu m\sum_{n=0}^{+\infty}F_{2n,1}\zeta^{2n}-\frac{i}{2}m\bigg(u^\mu (S\cdot q)^\nu+u^\nu (S\cdot q)^\mu\bigg)\sum_{n=0}^{+\infty}F_{2n+1,3}\zeta^{2n} .
\label{EMT_4D}
\end{equation}
In particular, one can characterize the multipolar structure of the Kerr solution \cite{Vines:2017hyw, Bianchi:2024shc}, which corresponds to
\begin{equation}
    F_{2\ell,1}=\frac{(-1)^\ell}{(2\ell)!} \qquad F_{2\ell+1,3}=\frac{(-1)^\ell}{(2\ell+1)!} \ .
\end{equation}
This leads to the well-known expression
\begin{equation}
    T^{\mu\nu}(q)\big|_{\text{Kerr}} =mu^\mu u^\nu\cos\zeta-\frac{i}{2}m\bigg(u^\mu(S\cdot q)^\nu+u^\nu(S\cdot q)^\mu\bigg)\frac{\sin\zeta}{\zeta} 
\end{equation}
for the stress-energy tensor of the Kerr solution in momentum space. Again, one can notice the similarities with  the exact metric in momentum space 
that can be computed in the Kerr-Schild gauge~\cite{Bianchi:2023lrg}.

\section{Multipoles from scattering amplitudes in four dimensions}
\label{sez 3}

In this section, we test the standard structure expected from the interaction between spinning matter and gravity. This analysis is useful for setting up the analogous computations in higher dimensions. In particular, the results obtained here will later be used for comparison with the specific case of five-dimensional spacetime.

As discussed in Section \ref{sez 2}, the expressions \eqref{multipoli} in four dimensions reduce to only two possible structures: the mass and the current multipole moments. 
The massive little group in four dimensions is $SO(3)\approx SU(2)$.
As shown in \eqref{Jirrep}, the multipolar expansion arises from decomposing the product of two polarizations into irreducible representations of the little group. In the case of $SU(2)$, the product of two irreducible \textbf{s} representations decomposes as
\begin{equation} \label{stimessgivesk}
    \mathbf{s}\otimes \mathbf{s}=\bigoplus_{k=0}^{2s}\mathbf{k} \ ,
\end{equation}
where the bold characters denote the representations, while the ordinary characters denote their labels. 
In terms of $SO(3)$, the irreducible representations of $SU(2)$ with integer $k$ correspond to completely symmetric and traceless tensors with $k$ vector indices. 
Therefore, a representation \textbf{k}, expressed in terms of the spin vector $s^i$, corresponds to an object of the form
\begin{equation}
    s^{i_1}s^{i_2}\ldots s^{i_{k-1}}s^{i_k} - {\rm traces}\ .
    \label{rappr k}
\end{equation}
This implies that the interaction of two spin-$s$ fields can generate all multipole moments up to the order $2s$.

We consider spin-1/2,  spin-1 and  spin-3/2 fields coupled to gravity, both minimally and non-minimally. This allows us to compute contributions up to the octupole order in the multipole expansion. In particular, the computations are expressed in term of the spin tensor $S^{ij}$ in order to maintain a uniform notation with the higher-dimensional case. Furthermore, we use the polarizations of the external massive fields such that, in the rest frame, their product is normalized as
\begin{equation}
    \varepsilon_1^A\cdot\varepsilon_{2,A}=1 \ .
\end{equation}
With this choice, the denominator in \eqref{SO(d) little group} is equal to one. The same convention will also be adopted in higher-dimensional spacetime.

\subsection{Spin 1/2 and universality of the dipole}
First, we analyze the minimal coupling between two massive spinors of equal mass $m$ and gravity. The corresponding amplitude produces only a monopole and a dipole term \cite{Bjerrum-Bohr:2002fji}, as it is obvious from the decomposition
\begin{equation}
    \mathbf{1/2\otimes1/2=1\oplus0} \ .
\end{equation}
The coupling of the spinor to gravity is described by the action
\begin{equation}
    S=\int  d^4x \, e\bar{\psi}(ie^\mu_a\gamma^a D_\mu-m)\psi \ ,
\end{equation}
where $e^\mu_a$ denotes the tetrad and the Dirac matrices are adopted with the following representation:
\begin{equation}
    \gamma^0=\begin{pmatrix}
        I_{2\times2}&0\\
        0& I_{2 \times 2}
    \end{pmatrix}\quad\quad \gamma^i=\begin{pmatrix}
        0& \sigma^i\\
        -\sigma^i&0
    \end{pmatrix} \ .
\end{equation}
 The Lorentz generators in the spinor representation are given by
\begin{equation}
    (M^{\rho\sigma})_\alpha{}^{\beta}=\frac{i}{2}\gamma^{\rho\sigma}\delta_\alpha^\beta \ ,
\end{equation}
where $\gamma^{\mu\nu}=1/2(\gamma^\mu\gamma^\nu-\gamma^\nu\gamma^\mu)$ and $\alpha,\beta$ are Dirac indices.
Thus, we obtain the stress-energy tensor
\begin{equation}
    T^{\mu\nu}_{h\psi^2}(q)=\bar{\varepsilon}_2(p_2)\bigg(\frac{1}{2}\gamma^\mu P^\nu+\frac{1}{2}\gamma^\nu P^\mu\bigg)\varepsilon_1(p_1) \ .
    \label{vert12}
\end{equation}
We now expand the spinor polarizations around the rest frame of each particle as
\begin{equation}
    \varepsilon_{1,2}(p_{1,2})=\varepsilon_{1,2}\mp \frac{\vec{\gamma}\cdot\vec{q}}{4m}\varepsilon_{1,2}+{\rm local \ terms}  \ ,
    \label{s1/2}
\end{equation}
where $\varepsilon_{1,2}$ are given by
\begin{equation}
    \varepsilon_{1,2}=\begin{pmatrix}
        \chi_{1,2}\\
        0
    \end{pmatrix}\ ,
\end{equation}
with $\chi_{1,2}$ being two-component non-relativistic spinors.
It is useful to note that
\begin{equation}
    \bar{\varepsilon}_2\gamma^0=\bar{\varepsilon}_2, \quad \gamma^0\varepsilon_1=\varepsilon_1 \ ,
\end{equation}
and 
\begin{equation}
    \bar{\varepsilon}_2\gamma^i\varepsilon_1=0 \ ,
\end{equation}
for $i=1,2,3$. 
To make the identification of the spin tensor explicit, we write the stress-energy tensor components:
\begin{equation}
    \begin{split}
&        T^{00}_{h\psi^2}(q)=m(\bar{\varepsilon}_2\varepsilon_1)\\
 &       T^{0i}_{h\psi^2}(q)=\frac{1}{2}\bigg[\bar{\varepsilon}_2\bigg(\frac{1}{2}i\epsilon^{ikl}\sigma_l\bigg)\varepsilon_1\bigg]q_k\\
  &      T^{ij}_{h\psi^2}(q)=0  \ , 
    \end{split} 
\end{equation}
Using the definition \eqref{SO(d) little group}, it is straightforward to identify the dipole term as 
\begin{align}
    \bigg[\bar{\varepsilon}_2\bigg(\frac{1}{2}i\epsilon^{ikl}\sigma_l\bigg)\varepsilon_1\bigg]=i\bar{\varepsilon}_2^{\, \alpha}(M^{ik})_\alpha{}^{\beta}\varepsilon_{1,\beta}=imS^{ik} \ .
\end{align}
This leads to 
\begin{equation}
\begin{split}
    & T^{00}_{h\psi^2}(q)=m\\
    & T^{0i}_{h\psi^2}(q)=\frac{1}{2}iS^{ik}q_k\\
    & T^{ij}_{h\psi^2}(q)=0 \ ,
\end{split} 
\end{equation}
which can be written in the covariant form
\begin{equation}
    T^{\mu\nu}_{h\psi^2}(q)=mu^\mu u^\nu-\frac{i}{2}mq_\rho(u^\mu S^{\nu\rho} +u^\nu S^{\mu\rho}) \ ,
\end{equation}
with $u^\mu=\delta^\mu_0$. 
This expression reproduces the metric of a vacuum rotating solution up to the dipole order \cite{Bjerrum-Bohr:2002fji}.
In particular, no additional operators can be added to the Lagrangian such that they contribute to the dipole. This is consistent with General Relativity, where the dipole term is universal for the metric of any vacuum solution in arbitrary dimension, and
an important consistency check of our approach is that the dipole coefficient generated by any spinning massive field is the same.

\subsection{Spin 1 and the quadrupole order}
In the case of a massive spin-1 field, \eqref{stimessgivesk} gives
\begin{equation}
    \bf{1}\otimes1=2\oplus1\oplus0 \ ,
\end{equation}
implying that the amplitude produces angular momentum terms up to the quadrupole order.
The minimal action is given by
\begin{equation}
    S=\int d^4x \, \sqrt{-g}\bigg(-\frac{1}{4}F_{\mu\nu}F^{\mu\nu}+\frac{1}{2}m^2A_\mu A^\mu\bigg) \ ,
\end{equation}
and the Lorentz generator in the vector representation is 
\begin{equation}
    (M^{\mu\nu})^{\rho\sigma}=i(\eta^{\mu\rho}\eta^{\nu\sigma}-\eta^{\mu\sigma}\eta^{\nu\rho}) \ .
\end{equation}
The stress-energy tensor in momentum space \cite{Bjerrum-Bohr:2014lea} reads 
\begin{equation}
\begin{split}
    T^{\mu\nu}_{hA^2}(q)&=\frac{1}{2m}\bigg\{(\varepsilon_1(p_1)\cdot p_2)\bigg(p_1^\mu\varepsilon_2^{*\nu}(p_2)+p_1^\nu\varepsilon_2^{*\mu}(p_2)\bigg)\\&+(\varepsilon_2^*(p_2)\cdot p_1)\bigg(p_2^\mu\varepsilon_1^\nu(p_1)+p_2^\nu\varepsilon_1^\mu(p_1)\bigg)-\varepsilon_2^*(p_2)\cdot\varepsilon_1(p_1)\bigg(p_1^\mu p_2^\nu+p_2^\mu p_1^\nu\bigg)\bigg\} \ .
\end{split}
\label{T_A}
\end{equation}
The polarization vectors $\varepsilon_{1,2}^\mu(p_{1,2})$, expanded around the rest frame of each particle, takes the form
\begin{equation}\label{polarizationvector}
    \varepsilon_{1,2}^{\mu}(p_{1,2})=\varepsilon_{1,2}^{\mu}\pm\frac{1}{2m}u^\mu(\vec{\varepsilon}_{1,2}\cdot\vec{q}\, )+\frac{1}{8m^2}q^\mu (\vec{\varepsilon}_{1,2}\cdot \vec{q} \, )+{\rm local \ terms} \ ,
\end{equation}
so that the components of \eqref{T_A} become
\begin{equation}
\begin{split}
&    T^{00}_{hA^2}(q)=m(\vec{\varepsilon}^{\ *}_2\cdot\vec{\varepsilon}_1)-\frac{1}{2m}(\vec{\varepsilon}_2^{\ *}\cdot\vec{q}\, )(\vec{\varepsilon}_1\cdot\vec{q}\, )\\
 &   T^{0i}_{hA^2}(q)=\frac{1}{2}(\varepsilon_1^k\varepsilon_2^{*i}-\varepsilon_1^i\varepsilon_2^{*k})q_k\\
 &   T^{ij}_{hA^2}(q)=0 \ ,
\end{split} 
\label{T_A components}
\end{equation}
Thus, from \eqref{SO(d) little group}, the dipole reads
\begin{equation}
    \epsilon_2^{i*}\epsilon_1^k-\epsilon_2^{k*}\epsilon_1^{i}=i\epsilon_2^{n*}(M^{ik})_n{}^{m}\epsilon_{1,m} = i m S^{ik}\ .
\end{equation}
Therefore, \eqref{T_A components} becomes
\begin{equation}
    \begin{split}
&        T^{00}_{hA^2}(q)=m+\frac{m}{2}(q_iS^{ik}S_{k}{}^{l} q_l)\\
 &       T^{0i}_{hA^2}(q)=\frac{i}{2}mS^{ik}q_k\\
  &      T^{ij}_{hA^2}(q)=0 \ ,
    \end{split} 
\end{equation}
which can be expressed in Lorentz covariant form as
\begin{equation}
    T^{\mu\nu}_{hA^2}(q)=u^\mu u^\nu m\bigg(1-\frac{1}{2}(q\cdot S\cdot S\cdot q)\bigg)-\frac{i}{2}mq_\rho(u^\mu S^{\nu\rho}+u^\nu S^{\mu\rho}) \ .
\end{equation}
This stress-energy tensor gives rise to the Kerr solution up to the quadrupole order in the multipole expansion.

In order to modify the quadrupole structure \cite{Vaidya:2014kza}, the only possible non-minimal operator at the tree level is
\begin{equation}
    \Delta\mathcal{L}_2^A=\frac{C_2^A}{4m^2}R_{\mu\nu\rho\sigma}F^{\mu\nu}F^{\rho\sigma} \ ,
    \label{op vec}
\end{equation}
with $C_2^A$ adimensional coupling.
This operator induces the following correction to the stress-energy tensor:
\begin{equation}
    T^{\mu\nu}_{hA^2}(q)=u^\mu u^\nu m\bigg(1-\bigg(\frac{C_2^A}{2}+\frac{1}{2}\bigg)(q\cdot S\cdot S\cdot q)\bigg)-\frac{i}{2}mq_\rho(u^\mu S^{\nu\rho}+u^\nu S^{\mu\rho}) \ .
\end{equation}
Comparing  this expression to the stress-energy tensor in \eqref{EMT_4D}, we can identify
\begin{equation}
    F_{2,1}=-\frac{1}{2}(C_2^A+1) \ .
\end{equation}
This gives rise to the most-general spin-induced vacuum solution up to quadratic order in the spin \cite{Gambino:2024uge}.

\subsection{Spin 3/2 and the octupole order}
For massive spin-3/2 fields, the amplitude generates multipoles up to the octupole order~\cite{Gherghetta:2024tob}, as can be seen from the decomposition
\begin{equation}
    \mathbf{3/2\otimes3/2=3\oplus2\oplus1\oplus0} \ .
\end{equation}
The minimal action is the  massive Rarita-Schwinger action,
\begin{equation}
    S=\int d^4x \, \sqrt{-g}\bigg(-\bar{\chi}_\mu(\epsilon^{\mu\rho\sigma\nu}\gamma_5\gamma_\rho D_\sigma+m\gamma^{\mu\nu})\chi_\nu\bigg) \ .
\label{Action RS}
\end{equation}
For the polarization of the spin-3/2 field, the following constraints hold:
\begin{equation}
    \begin{split}
        &p_{1,2}^\mu \varepsilon_{\mu,1,2}(p_{1,2})=0 \ ,\\&
        \gamma_\mu\varepsilon_{1,2}^\mu(p_{1,2})=0 \ , \quad \bar{\varepsilon}_{1,2}^\mu(p_{1,2})\gamma_\mu=0 \ .
    \end{split}
\end{equation}
In addition, the Dirac equation is satisfied:
\begin{equation}
    (\slashed{p}_{1,2}-m)\varepsilon_{1,2}^\mu(p_{1,2})=0 \ , \quad \bar{\varepsilon}_{1,2}^\mu(p_{1,2})(\slashed{p}_{1,2}-m)=0 \ .
\end{equation}
The Lorentz generator in this representation is given by
\begin{equation}
    (M^{\rho\sigma})^{\mu}{}_{\alpha}{}^{\nu \beta}=i\bigg(\frac{1}{2}(\gamma^{\rho\sigma})_{\alpha}{}^{\beta} \eta^{\mu\nu}+(\eta^{\rho\mu}\eta^{\sigma\nu}-\eta^{\rho\nu}\eta^{\sigma\mu})\delta_\alpha^\beta\bigg) \ .
\end{equation}
The polarization expanded around the rest frame of each particle takes the form
\begin{equation}
\begin{split}
    \varepsilon_{1,2}^\mu(p_{1,2})&= \varepsilon^\mu_{1,2}\mp\frac{1}{4m}(\vec{q}\cdot\vec{\gamma})\varepsilon_{1,2}^\mu\pm\frac{u^\mu}{2m}(\vec{q}\cdot\vec{\varepsilon}_{1,2})-\frac{u^\mu}{8m^2}(\vec{q}\cdot\vec{\gamma})(\vec{q}\cdot\vec{\varepsilon}_{1,2})\\&+
    \frac{1}{8m^2}q^\mu(\vec{q}\cdot\vec{\varepsilon}_{1,2})\mp\frac{1}{32m^3}q^\mu(\vec{q}\cdot\vec{\gamma})(\vec{q}\cdot\vec{\varepsilon}_{1,2})+ {\rm local \ terms} \ .
\end{split}
\end{equation}
The resulting stress-energy tensor from \eqref{Action RS} is
\begin{equation}
\begin{split}
    T^{\mu\nu}_{h\chi^2}(q)=&-\frac{1}{2}\bar{\varepsilon}_{2,\rho}(p_2)(P^\mu\gamma^\nu+P^\nu\gamma^\mu)\varepsilon^\rho_1(p_1)+\frac{1}{2}(\bar{\varepsilon}_2(p_2)\cdot p_1)(\gamma^\mu \varepsilon^\nu_1(p_1)+\gamma^\nu \varepsilon_1^\mu(p_1))\\&+\frac{1}{2}(\bar{\varepsilon}_2^{\ \nu}(p_2)\gamma^\mu+\bar{\varepsilon}_2^{\ \mu}(p_2)\gamma^\nu)(\varepsilon_1(p_1)\cdot p_2) \ .
\end{split}
\label{T3/2}
\end{equation}
Expanding around the rest frame, the components of \eqref{T3/2} become
\begin{align}
    \begin{split}
     &   T^{00}_{h\chi^2}(q)=m(\vec{\bar{\varepsilon}}_2\cdot \vec{\varepsilon}_1)-\frac{(\vec{q}\cdot\vec{\bar{\varepsilon}}_2)(\vec{q}\cdot\vec{\varepsilon}_1)}{2m}\\
      &  T^{0i}_{h\chi^2}(q)=\frac{1}{2}\bigg(\bar{\varepsilon}_{2,l}\bigg(i\frac{\epsilon^{ink}\sigma_k}{2}\bigg)\varepsilon_{1,l}+(\bar{\varepsilon}_2^{\ i}\varepsilon_1^n-\bar{\varepsilon}_2^{\ n}\varepsilon_1^i)\bigg)q_n-\frac{1}{8}\bigg(\bar{\varepsilon}_{2,l}(i\epsilon^{ink}\sigma_k)\varepsilon_1^m\bigg)q_nq_mq_l\\
       & T^{ij}_{h\chi^2}(q)=\frac{1}{4m}\bigg(\bar{\varepsilon}_2^{\ i}(i\epsilon^{jnk}\sigma_k)\varepsilon_1^l-\bar{\varepsilon}_2^{\ l}(i\epsilon^{jnk}\sigma_k)\varepsilon_1^i\bigg)q_nq_l+(i\leftrightarrow j) \ .
    \end{split} 
\end{align}
Using \eqref{SO(d) little group}, the dipole term reads
\begin{equation}
    \bigg[\bar{\varepsilon}_{2,n}\bigg(-\frac{1}{2}i\epsilon^{ilk}\sigma_k\bigg)\varepsilon_{1}^n+(\bar{\varepsilon}_2^{\ i}\varepsilon_1^l-\bar{\varepsilon}_2^{\ l}\varepsilon_1^i)\bigg]=i\epsilon^{ilk}\bar{\varepsilon}_2^{\ n\alpha}(M^k)_{n\alpha}{}^{ m\beta}\varepsilon_{1,m\beta} = i m S^{il} \ ,
\end{equation}
where $\alpha,\beta$ denote Dirac indices, while the latin indices refer to Euclidean indices.
For the octopole computation, the following identity is useful:
\begin{equation}
    \epsilon^{inb}(\vec{\sigma}\cdot\vec{q}\, )q_n=q^b\epsilon^{ink}\sigma_kq_n+{\rm local \ terms} \ .
\end{equation}
Thus, the stress-energy tensor in Lorentz covariant form reads
\textbf{\begin{equation}
    T^{\mu\nu}_{h\chi^2}(q)=u^\mu u^\nu m\bigg(1-\frac{1}{2}(q\cdot S\cdot S\cdot q)\bigg)-\frac{i}{2}mq_\rho(u^\mu S^{\nu\rho}+u^\nu S^{\mu\rho})(1-\frac{1}{6}(q\cdot S\cdot S\cdot q)) \ .
\label{T_RS}
\end{equation}}
This stress-energy tensor reproduces the Kerr solution up to octopole order.

To generalize \eqref{T_RS}, we introduce,
for the quadrupole and the octopole terms respectively, the non-minimal operators
\begin{equation}
\begin{split}
    &\Delta\mathcal{L}_2^\chi=\frac{C_2^\chi}{m^3} R_{\mu\nu\rho\sigma}D^\rho\bar{\chi}^\mu D^\nu\chi^\sigma \ , \\&
    \Delta\mathcal{L}_3^\chi=\frac{C_3^\chi}{4m^3}R_{\mu\nu\rho\sigma}D^\mu\bar{\chi}^\nu(\gamma^{\rho\tau}D_\tau)\chi^\sigma \ ,
\end{split}
\end{equation}
with $C_2^\chi$ and $C_3^\chi$ adimensional couplings.
They induce the following general stress-energy tensor:
\begin{equation}
\begin{split}
    T^{\mu\nu}_{h\chi^2}(q)&=u^\mu u^\nu m\bigg(1+\bigg(\frac{C_2^\chi}{2}-\frac{1}{2}\bigg)(q\cdot S\cdot S\cdot q)\bigg)\\&-\frac{i}{2}mq_\rho(u^\mu S^{\nu\rho}+u^\nu S^{\mu\rho})\bigg(\frac{1}{6}(C_3^\chi-1)(q\cdot S\cdot S\cdot q)\bigg) \ .
\end{split}
\end{equation}
By comparison with \eqref{EMT_4D}, we identify
\begin{equation}
    F_{2,1}=\frac{1}{2}(C_2^\chi-1), \quad F_{3,3}=\frac{1}{6}(C_3^\chi-1) \ ,
\end{equation}
and the corresponding metric is the most-general spin-induced vacuum solution up to octupole order.

\section{Multipoles from scattering amplitudes in five dimensions}
\label{sez 4}
We now consider the higher-dimensional case, focusing for simplicity on five dimensions. In this case, the massive little group is $SO(4)\approx SU(2)_L\otimes SU(2)_R$. Therefore, there are two quantum little group numbers, allowing for a richer set of field representations.
Besides, the stress moments in \eqref{multipoli} no longer vanish.

The interaction between two representation ($\mathbf{s}_1$,$\mathbf{s}_2$) decomposes as
\begin{equation}\label{productofspinsD=5}
(\mathbf{s}_1,\mathbf{s}_2)\otimes(\mathbf{s}_1,\mathbf{s}_2)=\bigoplus_{k_1=0}^{2s_1}\bigoplus_{k_2=0}^{2s_2}(\mathbf{k}_1,\mathbf{k}_2) \ ,
\end{equation}
with integer $k_1$ and $k_2$.
Among the representations that appear on the right-hand side of \eqref{productofspinsD=5}, we expect that only the ones  that correspond to the multipoles in \eqref{STFASTFRSTF} will survive in the scattering amplitude in the classical limit. 
In particular, the representations with $k_1=k_2$ correspond to $2k_1$ totally symmetric traceless (STF) $SO(4)$ vector indices, giving rise to the mass multipoles.  
The representations with $k_1 \neq k_2$  correspond to  mixed-symmetry objects. In particular, the $(\mathbf{k},\mathbf{k -1})\oplus(\mathbf{k}-1,\mathbf{k}) $ representation corresponds to $2k$ indices of $SO(4)$ in the ASTF representation (current multipoles), while  the $(\mathbf{k+1},\mathbf{k -1})\oplus(\mathbf{k}-1,\mathbf{k+1}) $ representation corresponds to $2k+2$ indices of $SO(4)$ in the RSTF representation (stress multipoles). This allows us to identify the multipoles as
the following representations:
\begin{equation}
\begin{split}
    2\ell{\rm -order \ mass \ multipoles:} \qquad & (\mathbf{\ell}, \mathbf{\ell})\\
    (2\ell+1){\rm -order \ current \ multipoles:} \qquad & (\mathbf{\ell+1}, \mathbf{\ell})\oplus (\mathbf{\ell}, \mathbf{\ell+1})\\ 
    2\ell{\rm -order \ stress \ multipoles:} \qquad &(\mathbf{\ell+1}, \mathbf{\ell-1})\oplus (\mathbf{\ell-1}, \mathbf{\ell+1}) \ .    
\end{split}
\label{multipolesreprd=5}
\end{equation}
All the other representations produced in \eqref{productofspinsD=5} do not correspond to multipoles, and therefore by consistency we expect them not to appear in the scattering amplitude in the classical limit, because they vanish when contracted with the momenta up to local or gauge terms. Verifying that this happens in an actual scattering computation is an important consistency check. 

In this section, we analyze a vector field and an antisymmetric tensor. We will see that at the quadratic order in the angular momentum, the vector field can only produce a mass quadrupole, while the antisymmetric tensor only produces a stress quadrupole term. 

\subsection{Vector}
In five dimensions, the vector polarization belongs to the representation (\textbf{1/2},\textbf{1/2}) of the little group. The interaction between two vectors decomposes as
\begin{equation}
    \bf{(1/2,1/2)}\otimes(1/2,1/2)=(1,1)\oplus(1,0)\oplus(0,1)\oplus(0,0) \ .
\end{equation}
From \eqref{multipolesreprd=5}, the representation (\textbf{1},\textbf{1}) the mass quadrupole, the representation  (\textbf{1},\textbf{0})$\oplus$(\textbf{0},\textbf{1}) corresponds to the dipole, and the representation (\textbf{0},\textbf{0}) corresponds to the monopole \cite{Gambino:2024uge}.
From this decomposition, we can immediately see that the stress moment cannot be generated by the interaction of two vector fields. We will confirm this explicitly in the calculations below.

The spin operator is the same as the four dimensions, and the polarization vector is expanded around the rest frame as in \eqref{polarizationvector}, but now there are four spatial dimensions. This leads to a different coefficient in the quadrupole term.
Specifically, for the quadrupole moment, we have
\begin{equation}
        q_lq_n(S^{lk}S_k{}^{n})\big|_{\rm STF}=-\frac{2}{m^2}(\varepsilon_2^{l*}q_l)(\varepsilon_1^{n}q_n) \ .
\end{equation}
\\
Thus, the covariant form of the minimal stress-energy tensor takes the form
\begin{equation}
    T^{\mu\nu}_{hA^2}(q)\bigg|_{d=4}=u^\mu u^\nu m\bigg(1-\frac{1}{4}(q\cdot S\cdot S\cdot q)\bigg)-\frac{i}{2}mq_\rho(u^\mu S^{\nu\rho}+u^\nu S^{\mu\rho}) \
    .
    \label{T1}
\end{equation}
As we can see, the vector field in different dimensions produces a different quadrupole term, demonstrating that the spin interaction is not universal in arbitrary dimensions. 
If we attempt to generate the stress quadrupole, we find
\begin{equation}
    q_kq_l(S^{ik}S^{jl})\big|_{\rm RSTF}=0 \ ,
\end{equation}
which is consistent with the group theory argument that a vector field cannot produce a stress quadrupole. Therefore, there is no operator that we can add to the Lagrangian such that it generates a stress quadrupole term. The operator that can modify the mass quadrupole is again \eqref{op vec}.
In particular, the stress-energy tensor \eqref{T1} does not reproduce the mass quadrupole of Myers Perry solution.

One can generalize to an arbitrary dimension $D=d+1$. The quadrupole moments are
\begin{equation}
    \begin{split}
        &q_lq_n(S^{lk}S_k{}^{n})\big|_{\rm STF}=-\frac{1}{m^2}(d-2)(\varepsilon_2^{l*}q_l)(\varepsilon_1^{n}q_n) \ ,\\&
        q_kq_l(S^{ik}S^{jl})\big|_{\rm RSTF}=0 \ .
    \end{split}
\end{equation}
This means that a general feature of the vector field in arbitrary dimensions is that it can generate only the mass quadrupole and not the stress quadrupole.
The Lorentz covariant form of minimal $T^{\mu\nu}$ in arbitrary dimensions is
\begin{equation}
    T^{\mu\nu}_{hA^2}(q)\bigg|_{D=d+1}=u^\mu u^\nu m\bigg(1-\frac{1}{2(d-2)}(q\cdot S\cdot S\cdot q)\bigg)-\frac{i}{2}mq_\rho(u^\mu S^{\nu\rho}+u^\nu S^{\mu\rho}) \ .
\end{equation}

\subsection{Anti-symmetric tensor}
In five-dimensional spacetime, a massive antisymmetric tensor transforms in the reducible representation of $SU(2)_L\otimes SU(2)_R$
\begin{equation}
    \bf(1,0)\oplus(0,1) \ .
\end{equation}
The direct sum of the two irreducible representations is realized in terms of sum of a self-dual and an antiself-dual tensors:
\begin{equation}
    B^{ij}=(B^{+,ij}+B^{-,ij}) \ ,
\end{equation}
with
\begin{equation}
    B^{+,ij}=\frac{1}{2}\epsilon^{ijkl}B^{+}_{kl} \ , \quad B^{-,ij}=-\frac{1}{2}\epsilon^{ijkl}B^-_{kl} \ .
\end{equation}

The interaction between two antisymmetric tensors decomposes as
\begin{equation}
    {[(\mathbf{1},\mathbf{0})\oplus(\mathbf{0},\mathbf{1})]}\otimes[(\mathbf{1},\mathbf{0})\oplus(\mathbf{0},\mathbf{1})]=[(\mathbf{2},\mathbf{0})\oplus(\mathbf{0},\mathbf{2})]\oplus2\times(\mathbf{1},\mathbf{1})\oplus[(\mathbf{1},\mathbf{0})\oplus(\mathbf{0},\mathbf{1})]\oplus2\times(\mathbf{0},\mathbf{0}) \ .
\end{equation}
From \eqref{multipolesreprd=5}, the representation (\textbf{1},\textbf{0})$\oplus$(\textbf{0},\textbf{1}) corresponds to the dipole, and the representation (\textbf{2},\textbf{0})$\oplus$(\textbf{0},\textbf{2}) corresponds to the stress quadrupole. The two (\textbf{0},\textbf{0})s are the monopoles generated by the product of two self-dual tensors and of two antiself-dual tensors. 
One might expect that (a combination of) the two (\textbf{1},\textbf{1})s correspond to the mass quadrupole arising from the product of a self-dual tensor and an antiself-dual tensor. However, considering the action of a spin operator between states belonging to different irreducible representations, one finds that it vanishes:
\begin{equation}
    {\langle}l_2,m_2|S^n|l_1,m_1\rangle=0 \ .
\end{equation}
Therefore, as will be shown explicitly through direct computation, the corresponding mass quadrupole vanishes.

For the anti-symmetric field, the Lorentz generator is given by
\begin{equation}
\begin{split}
    (M^{\mu\nu})^{\rho\sigma ,\tau\lambda}=\frac{i}{2}(&+\eta^{\rho\tau}\eta^{\lambda\nu}\eta^{\mu\sigma}-\eta^{\rho\tau}\eta^{\lambda\mu}\eta^{\nu\sigma}+\eta^{\lambda\sigma}\eta^{\tau\nu}\eta^{\mu\rho}-\eta^{\lambda\sigma}\eta^{\tau\mu}\eta^{\nu\rho}\\&
    -\eta^{\sigma\tau}\eta^{\lambda\nu}\eta^{\mu\rho}+\eta^{\sigma\tau}\eta^{\lambda\mu}\eta^{\nu\rho}-\eta^{\lambda\rho}\eta^{\tau\nu}\eta^{\mu\sigma}+\eta^{\lambda\rho}\eta^{\tau\mu}\eta^{\nu\sigma}) \ .
\end{split}
\end{equation}
The expansion of the polarization $\varepsilon^{\mu\nu} (p)$  around the rest frame polarization tensor $\varepsilon^{\mu\nu}$ reads 
\begin{align}
\begin{split}
    \varepsilon_{1,2}^{\mu\nu}(p_{1,2})& = \varepsilon^{\mu\nu}_{1,2}\pm\frac{1}{2m}\bigg(u^\nu(\varepsilon^\mu{}_{1,2,i}q^i)-u^\mu(\varepsilon^\nu{}_{1,2,i}q^i)\bigg)\\&+\frac{1}{8m^2}\bigg(q^\nu(\varepsilon^\mu{}_{1,2, i}q^i)-q^\mu(\varepsilon^\nu{}_{1,2, i}q^i)\bigg)+{\rm local \ terms} \ 
\end{split}
\end{align}
with $\varepsilon^{\mu0}_{1,2}=0$.

The minimal coupling between an antisymmetric tensor $B^{\mu\nu}$ and gravity is given by the action
\begin{equation}
    S=\int d^5x \, \sqrt{-g}\bigg(-\frac{1}{6}H_{\mu\nu\rho}H^{\mu\nu\rho}-\frac{1}{2}m^2B_{\mu\nu}B^{\mu\nu}\bigg) \ ,
\label{Action B}
\end{equation}
where the field strength $H_{\mu\nu\rho}$ is defined as
\begin{equation}
    H_{\mu\nu\rho}=\partial_\mu B_{\nu\rho}+\partial_\rho B_{\mu\nu}+\partial_\nu B_{\rho\mu} \ .
\end{equation}
The stress-energy tensor arising from \eqref{Action B} is
\begin{equation}
\begin{split}
    T^{\mu\nu}_{hB^2}(q)=\frac{1}{2m}\bigg\{&-\bigg(p_1^\mu p_2^\nu+p_1^\nu p_2^\mu\bigg)\varepsilon_{1,\rho\sigma}(p_1 )\varepsilon^{*,\rho\sigma}_2(p_2 )\\&-2\bigg(\varepsilon_{1,\rho\sigma}(p_1)p_2^\rho\bigg)\bigg(p_1^\mu\varepsilon^{*,\sigma\nu}_2(p_2)+p_1^\nu\varepsilon_2^{*,\sigma\mu}(p_2)\bigg)\\&
    -2\bigg(\varepsilon^*_{2,\rho\sigma}(p_2)p_1^\rho\bigg)\bigg(p_2^\mu\varepsilon_1^{\sigma\nu}(p_1)+p_2^\nu\varepsilon_1^{\sigma\mu}(p_1)\bigg)\\&-2p_1^\rho p_2^\sigma\bigg(\varepsilon_1^{\mu}{}_{\sigma}(p_1)\varepsilon_{2,\rho}^{*}{}^\mu(p_2)+\varepsilon^{\nu}_{1}{}_{\sigma}(p_1)\varepsilon_{2,\rho}^{*}{}^{\mu}(p_2)\bigg)\bigg\} \ .
\end{split}
\label{TB}
\end{equation}
Expanding the polarizations in \eqref{TB} in terms of the rest-frame polarizations one gets
\begin{equation}
    \begin{split}
&        T^{00}_{hB^2}(q)=m(\varepsilon_2^*{}_{,ij}\ \varepsilon_1^{ij})+\frac{1}{2m}q_mq_n(\varepsilon_{2}^{*,n}{}_{k}\varepsilon_1^{km}+\varepsilon_{2}^{*,m}{}_{k}\varepsilon_1^{kn})\\
 &       T^{0i}_{hB^2}(q)=-q_k(\varepsilon_2^{*,il}\varepsilon_{1,l}{}^{k}-\varepsilon_2^{*,kl}\varepsilon_{1,l}{}^{i})\\
  &      T^{ij}_{hB^2}(q)=\frac{1}{m}q_n q_m(\varepsilon_2^{*,jm}\varepsilon_1^{jn}+\varepsilon_2^{*,jn}\varepsilon_1^{jm})\ .
    \end{split} 
\label{T_B components}
\end{equation}
Using \eqref{SO(d) little group}, the dipole is identified as
\begin{equation}
\varepsilon_2^{*,ik}\varepsilon_1^{kj}-\varepsilon_2^{*,jk}\varepsilon_1^{ki}=-\frac{i}{2}\varepsilon_2^{*,ab}(M^{ij})_{ab}{}^{cd}\varepsilon_{1,cd} =-\frac{i}{2} m S^{ij} \ .
\label{Spinantisymmetrictensor}    
\end{equation}
Computing the STF quadrupole structure, using \eqref{Jirrep}, we find
\begin{equation}
    q_nq_l(S^{nk}S_k{}^{l})\big|_{\rm STF}=0 \ .
\end{equation}
As we can see, in five dimensions the mass quadrupole term vanishes. Therefore, it does not contribute to the multipole expansion.
The vanishing STF structure implies that the product of the self-dual and the anti-self dual tensors does not yield a classical contribution. This is consistent with the initial consideration of this section. Consequently, the term  $q\cdot\varepsilon_2^\dagger\cdot\varepsilon_1\cdot q$ is to be regarded only as a product of either two self-dual or two antiself-dual tensors. However, such products are proportional to the singlets and thus are local
\begin{equation}
\begin{split}
    &q\cdot\varepsilon_2^{+,*}\cdot\varepsilon_1^+\cdot q= q^2\varepsilon_2^{+,*}\cdot\varepsilon_1^+ \ , \\&
    q\cdot\varepsilon_2^{-,*}\cdot\varepsilon_1^-\cdot q= q^2\varepsilon_2^{-,*}\cdot\varepsilon_1^- \ .
\end{split}
\end{equation}
Therefore, these contributions can be neglected.

We can now compute the RSTF quadrupole structure, which is
\begin{equation}
     q_kq_l(S^{ik}S^{jl})\big|_{\rm RSTF}=-\frac{2}{m^2}(\varepsilon_2^{*, \ ik}\varepsilon_1^{jl}+\varepsilon_2^{*, \ jl}\varepsilon_1^{ik})q_lq_k \ .
\end{equation}
This implies that, unlike the vector case, in five dimensions the antisymmetric tensor contributes, either minimally or non-minimally, only to the stress quadrupole.
Finally, the stress-energy tensor \eqref{T_B components} in Lorentz covariant form reads
\begin{equation}
    T^{\mu\nu}_{hB^2}(q)=u^\mu u^\nu m-\frac{i}{2}mq_\rho(u^\mu S^{\nu\rho}+u^\nu S^{\mu\rho})-\frac{m}{2}(S \cdot q)^\mu(S\cdot q)^\nu \ .\label{T_Bminimal}
\end{equation}

Also in this case, the minimal coupling does not reproduce the stress quadrupole structure of the Myers-Perry solution.
To modify the stress quadrupole, one can add to the Lagrangian the operator
\begin{equation}
    \Delta\mathcal{L}^B_2=\frac{C_{2,2}^B}{m^2}R_{\mu\nu\rho\sigma}H^{\mu\nu\tau}H_\tau^{\ \rho\sigma}  \ ,
\end{equation}
with $C_{2,2}^B$ adimensional coupling.
This leads to the most general stress-energy tensor
\begin{equation}
    T^{\mu\nu}_{hB^2}(q)=mu^\mu u^\nu-\frac{i}{2}mq_\rho\bigg(u^\mu S^{\nu\rho}+u^\nu S^{\mu\rho}\bigg)+m\bigg(\frac{C_{2,2}^B}{2}-\frac{1}{2}\bigg)(S\cdot q)^\mu(S\cdot q)^\nu \ .
\label{T_B}
\end{equation}
From \eqref{EMT}, we have the correspondence
\begin{equation}
    F_{2,2}=\frac{1}{2}(C_{2,2}^B-1) \ .
\end{equation}

To conclude this section, we observe that the vanishing of the mass quadrupole in the scattering of  a massive antisymmetric tensor is a unique feature of the five dimensional case.
Indeed, considering the antisymmetric tensor in arbitrary dimensions $D=d+1$, from \eqref{Spinantisymmetrictensor} it follows that the mass quadrupole is
\begin{equation}
    q_nq_l(S^{nk}S_k{}^{l})\big|_{\rm STF}=\frac{2}{m^2}(d-4)q_k\varepsilon_2^{*,kl}\varepsilon_{1,l}{}^{n}q_n \ .
\end{equation}
Therefore, in $d>4$ the mass quadrupole is no longer vanishing, so it can be generated by the antisymmetric tensor. This is consistent with the fact that in higher dimensions the decomposition in self-dual and antiself-dual is no longer valid. Nonetheless, we note that the minimal stress-energy tensor is equal to \eqref{T_Bminimal} for any spacetime dimensions and in particular, using \eqref{Duality S tensor}, one can notice that in $d=3$ it reproduces the multipolar structure of the Kerr solution up to the quadrupole.

\section{Conclusions}
\label{sez 5}
In this paper we have studied the multipolar structure of vacuum rotating solutions in general relativity in higher dimensions from a scattering-amplitude perspective. While in four dimensions the 3-point amplitude describing the emission of a graviton by a massive spin $s$ field  produces all terms in the multipolar expansion up to $2s$ order \cite{Arkani-Hamed:2017jhn}, in higher dimensions 
this universality is lost. We focused in particular on the five-dimensional case, and we studied the quadrupole term produced by a massive vector and by a massive antisymmetric tensor. We showed that while the former generates only a mass quadrupole, exactly the opposite applies for the latter, that produces only a stress quadrupole. This is a particular case of a completely general 
 loss of spin universality in the interaction between spinning matter and gravity in higher dimensions. This loss of spin universality once again highlights the uniqueness of four-dimensional spacetime, being related, as our analysis shows, to the fact that in higher dimensions the little group has rank higher than 1, and this leads to a much richer spectrum of representations. 
 Although our results are all tree-level, corresponding to the first order in the post-Minkowskian expansion, there is no major obstacle in extending them to higher orders by including loop corrections \cite{Donoghue:2001qc,Bjerrum-Bohr:2002fji,Holstein:2004dn, Bjerrum-Bohr:2018xdl, Damgaard:2026kqg}. As shown in \cite{Mougiakakos:2020laz, DOnofrio:2022cvn} in the non-rotating case and generalized in  \cite{Gambino:2024uge} for the case of spinning particles, in specific gauges and in specific dimensions loop diagrams can exhibit infrared divergences, which lead to singularities in the metric. This issue can be addressed by introducing higher-derivative couplings as counter-terms. 
 
 It would be of extreme interest to investigate whether spin universality can be restored in the infinite-spin limit, by performing a higher-spin analysis, in which all angular momenta go to infinity, along the lines of refs. \cite{Cangemi:2022bew, Cangemi:2023ysz}. 
This represents a natural direction for future investigation. In the specific case of five dimensions, the massive spinor-helicity formalism  developed by \cite{Chiodaroli:2022ssi, Pokraka:2024fao} should be employed to shed light in this direction. More generally, another line of investigation could be to explore the systematics of the higher-derivative terms that give rise to the Myers-Perry multipolar structure discussed in this paper,  in order to possibly find a string theory interpretation following the reasoning discussed in \cite{Alessio:2025nzd} for the Kerr case.\footnote{We thank M. Bianchi for pointing out this possibility to us.}

Another natural extension of this work is the study of the multipolar structure of  charged rotating  objects in higher dimensions, generalizing the results of \cite{Gambino:2025iyx}, where only the dipole order was considered. It this context, it would be interesting to connect the analysis with the electromagnetic multipolar structure of the five-dimensional solution of \cite{Chong:2005hr}. Recently, for specific values of the angular momenta such solution has been extended to odd dimensions higher than five \cite{Deshpande:2024vbn}, and generating this solution from scattering amplitudes would also be worth investigating.

Although extending four-dimensional results to higher dimensions does not have at first sight obvious phenomenological applications, in the study of the two-body problem in gravity this analysis might be relevant in the context of dimensional regularization  \cite{Akpinar:2025huz}. Shedding light on how the angular momentum expansion arises in higher dimensions would therefore be important in the context of studying binary systems at higher orders in the expansion of the spin.

\begin{acknowledgments}
We would like to thank Massimo Bianchi, Claudio Gambino and Paolo Pani for suggestions and for carefully reading the manuscript, as well as Dogan Akpinar, Lucile Cangemi and  Iustin Surubaru for valuable discussions. 
This work is partially supported by the INFN ST\&FI and TEONGRAV initiatives.
\end{acknowledgments}





\bibliographystyle{JHEP}
\bibliography{biblio.bib}

\end{document}